\pdfoutput=1
\documentclass[conference]{IEEEtran}
\IEEEoverridecommandlockouts
\usepackage{cite}
\usepackage[linesnumbered,ruled,vlined]{algorithm2e}
\usepackage{amsmath,amssymb,amsfonts}
\usepackage{algorithmic}
\usepackage{graphicx}
\usepackage{textcomp}
\usepackage{xcolor}
\usepackage{tikz}
\usepackage{array}
\usepackage{tabularx}
\usepackage{threeparttable}
\newcolumntype{Y}{>{\raggedleft\arraybackslash}X}
\usepackage{booktabs}
\usepackage{multirow}
\usepackage{makecell}
\usepackage{colortbl} 
\usepackage{subcaption} 

\newif\ifdraftnotes\draftnotesfalse
\definecolor{improvred}{RGB}{190,30,45}

\def\BibTeX{{\rm B\kern-.05em{\sc i\kern-.025em b}\kern-.08em
    T\kern-.1667em\lower.7ex\hbox{E}\kern-.125emX}}

\begin{document}
\bstctlcite{IEEEtranBSTCTL:allauthors}

\title{MoE Expert Execution in Disaggregated LLM Serving with a High-Bandwidth ReRAM Near-Memory Architecture}

\author{
\IEEEauthorblockN{Kunming Shao\textsuperscript{1,2,*},
Ming Zeng\textsuperscript{2},
Xin Yuan\textsuperscript{2},
Binbin Liao\textsuperscript{2},
Yangming Zhang\textsuperscript{2,*},}
\IEEEauthorblockN{Wei Wang\textsuperscript{2},
Tim Kwang-Ting Cheng\textsuperscript{1},
and Chi-Ying Tsui\textsuperscript{1}}
\IEEEauthorblockA{\textsuperscript{1}The Hong Kong University of Science and Technology, Hong Kong SAR, China\\
\textsuperscript{2}Alibaba Cloud Computing, China\\
Email: kshaoaa@connect.ust.hk, yangming.zym@alibaba-inc.com
\vspace{-3mm}
\thanks{*Corresponding authors: Kunming Shao and Yangming Zhang.}}
}

\maketitle

\begin{abstract}
Attention--FFN disaggregation maps LLM modules to specialized pools,
creating an opening to keep Mixture-of-Experts (MoE) weights resident in
a high-bandwidth FFN pool.
Decode SLOs, however, cap the run-batch while sparse routing expands the
activated-expert union, so weight traffic amortizes poorly and routing
skew idles cold-expert resources.
The FFN pool must therefore deliver weight-read bandwidth density under
sparse unions and recover occupancy under skew without a global sharing
fabric.
We present a ReRAM near-memory architecture that keeps expert
weights resident behind high-bandwidth local reads.
The design factors actual MFU into ideal MFU and occupancy, recovers
occupancy with bounded core-local multicast pooling, coactivation-aware
placement, and load-aware fetch, and sizes each communication level from
induced demand.
A measured + modeled study on Qwen3.5-35B-A3B, Qwen3.5-397B-A17B, and
GLM-5.2 shows that side-4 pooling raises occupancy from $0.328$ to
$0.519$ and, at iso-peak compute, lowers per-token FFN-pool latency by
$9.5\times$ versus H20 with $20\times$ lower weight-movement energy; an
H20-attention + ReRAM-FFN system reduces decode TPOT by
$1.25$--$4.0\times$, $2.4$--$10.3\times$, and $2.5$--$10.4\times$ versus
a homogeneous H20 pool.

\end{abstract}


\section{Introduction}
\label{sec:intro}

\begin{figure}[t]
\centering
\includegraphics[width=\columnwidth]{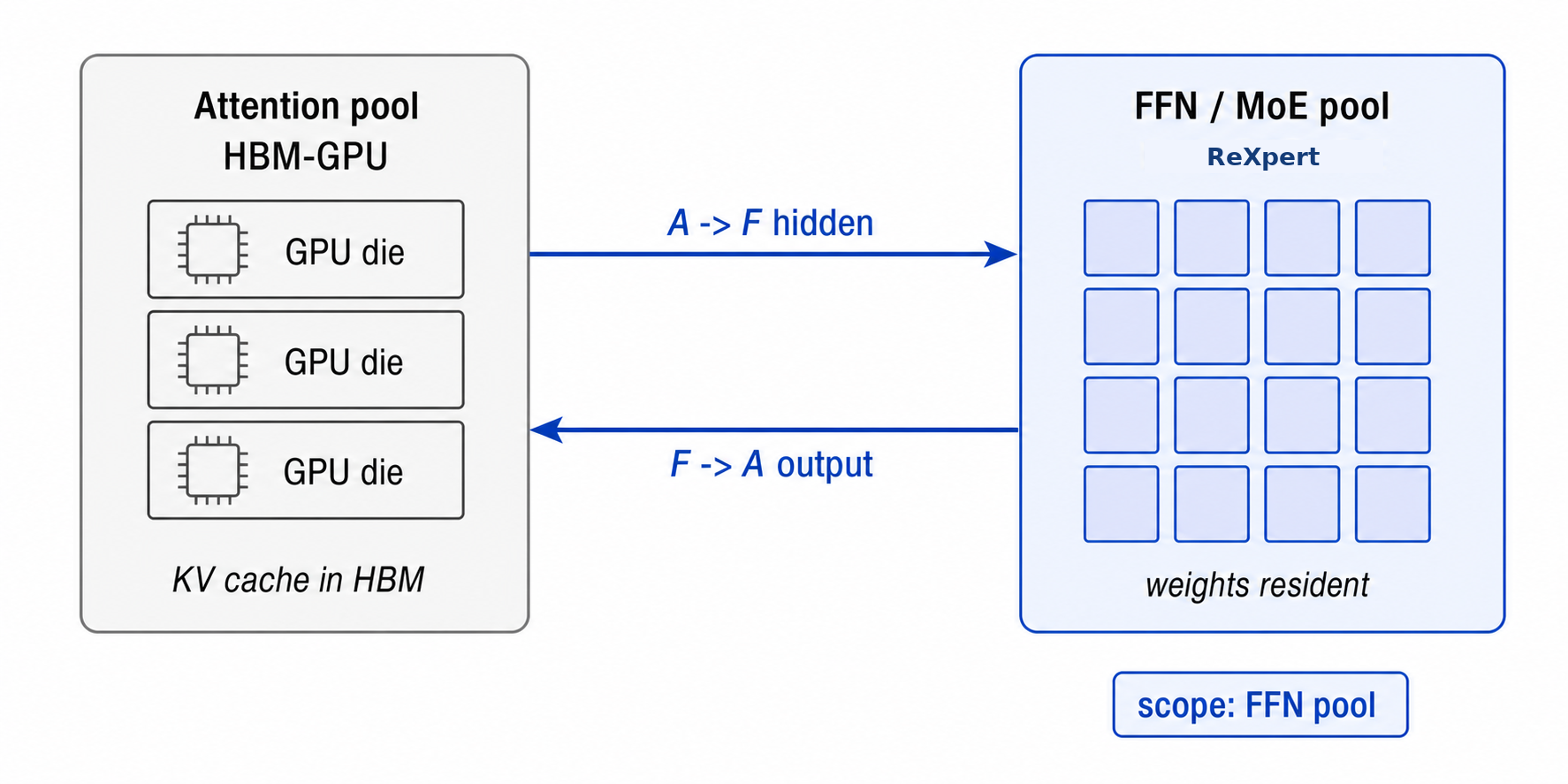}
\caption{Attention/FFN-disaggregated serving. The attention pool is KV-read
(bandwidth) bound; the FFN pool holds resident MoE expert weights and is the
scope of this paper.}
\label{fig:af_disagg}
\vspace{-2mm}
\end{figure}

Attention and feed-forward (FFN) layers impose different costs during LLM
decode. Attention repeatedly reads a growing key-value cache and is
bandwidth-bound; an FFN reuses model weights that are constant across tokens.
Attention--FFN disaggregation maps these modules to separate pools specialized
for their bottlenecks \cite{zhong2024distserve,patel2024splitwise}.
For Mixture-of-Experts (MoE) models, this split creates an opportunity to keep
expert weights resident in a high-bandwidth near-memory computing (NMC) pool
and stream activations through them (Fig.~\ref{fig:af_disagg}).

Realizing that opportunity faces a concrete challenge.
A decode service-level objective (SLO) limits how many sequences advance
together, so run-batch is capped rather than freely chosen.
Sparse routing then expands the \emph{union} of activated experts: on measured
Qwen3.5-35B-A3B traces the union grows from $8$ to $168.6$ of $256$ experts
between batch~$1$ and~$64$, so a $64\times$ batch increase cuts per-token
weight traffic by only ${\approx}3.0\times$.
Nonuniform routing further lets hot experts set the step latency while cold
experts' resources sit idle.
Together these effects define the paper's challenge: after residency removes
off-chip weight delivery, the MoE FFN pool must still (i)~provide
\emph{bandwidth density}---weight-read bandwidth per resident capacity---under
sparse unions, and (ii)~recover local utilization under skew without paying
for global expert replication or a global sharing fabric.

Prior MoE systems mainly cut weight or activation movement
\cite{hwang2024pregated,du2024sida,cao2025moelightning,kim2024monde,yun2024duplex}
or balance expert parallelism across GPUs
\cite{he2022fastermoe,li2023lina,zhai2023smartmoe}.
Once weights are resident, skew appears as two local losses: issued tiles can
be underfilled, and PE groups can wait on a straggler.
In-array compute would couple each PE to its programmed expert; reading a
resident row into adjacent digital PEs instead makes the destination PE a
runtime choice while keeping conventional FP8 datapaths.
ReXpert spends that local bandwidth for reassignability inside a physically
bounded neighborhood.

We present ReXpert, a ReRAM NMC architecture for the MoE FFN pool that combines
bounded core-local multicast pooling, routing-aware placement, and load-aware
fetch, and provisions each communication level from the induced traffic.

\begin{itemize}
\item \textbf{Challenge and bandwidth-density analysis.}
SLO-capped batch, sparse expert unions, and routing skew jointly motivate
resident capacity with proportionally high local read bandwidth.
\item \textbf{MFU decomposition and bounded pooling.}
Actual MFU $=$ ideal MFU $\times$ occupancy; multicast pooling and secondary
placement/fetch raise occupancy $0.328{\to}0.519$ without global sharing.
\item \textbf{Hierarchy-aware communication provisioning.}
Per-level traffic, placement-matched reduction order, and on-die headroom
for the induced demand.
\item \textbf{Hybrid evidence.}
Relative to attention-side AFD \cite{liu2026chime}, NPU--PIM splits
\cite{heo2024neupims,seo2024ianus}, and software MoE schedulers
\cite{cao2025moelightning,he2022fastermoe,yu2026patterns}, we keep routing
fixed and absorb skew inside a resident ReRAM NMC FFN pool.
At iso-peak compute the pool is $9.5\times$ faster than H20 with
$20\times$ lower weight-movement energy; an H20-attention~$+$~ReXpert-FFN
system lowers decode TPOT by $1.25$--$4.0\times$, $2.4$--$10.3\times$,
and $2.5$--$10.4\times$ on the three models versus a homogeneous H20 pool.
\end{itemize}

\section{Background and Motivation}
\label{sec:bg}

\subsection{Batch-Constrained AF Disaggregation}
\label{sec:bg-af}

Attention rereads the key-value (KV) cache for every decoded token, whereas an FFN
reuses weights that are constant across tokens. Disaggregated serving maps these
modules to separate pools provisioned for their resource profiles
\cite{zhong2024distserve,patel2024splitwise,qin2025mooncake}.

Decode SLOs constrain run-batch \cite{yu2022orca,kwon2023vllm}, so the FFN pool
must be efficient when latency forbids batching solely for weight amortization.
AFD also pipelines A/F waves and transfers activations between pools.
CHIME's Disaggregated Roofline develops the corresponding analysis for the
attention side \cite{liu2026chime}; we study only the MoE FFN side.

\subsection{Sparse-Expert Weight-Traffic Analysis}
\label{sec:bg-bw}

For a dense FFN, batching reduces per-token weight traffic as $1/B$
\cite{williams2009roofline,pope2023scaling}. A sparse MoE activates the union
$U(B)$ of the top-$r$ choices made by $B$ tokens, so traffic scales as $U(B)/B$.
On measured Qwen3.5-35B-A3B routing (default Alpaca trace), $U(B)$ grows from $8$
experts at $B{=}1$ to $168.6$ of $256$ at $B{=}64$ (and $205$ at native
$B{=}128$); a $64\times$ batch
increase therefore reduces
per-token traffic by only ${\approx}3.0\times$
($U(1)/1$ over $U(64)/64$).
This makes \emph{bandwidth density}---read
bandwidth per resident weight capacity---the relevant provisioning quantity.

\subsection{Technology-Independent Machine Balance}
\label{sec:bg-balance}

A machine with peak rate $P$ and weight-read bandwidth $\beta$ has balance
$F=P/\beta$ FLOP/byte. If $t$ tokens share each weight read, an
input-stationary GEMM performs $2t$ FLOP/byte and reaches the compute roof only
when
\begin{equation}
t \;\ge\; F/2
\label{eq:balance}
\end{equation}
Because mean expert load is $Br/R$, the corresponding batch is
$B^{\star}=(F/2)(R/r)$.
H20 has $F=74$ FLOP/byte and needs 37 tokens per read; H100/H800 has $F=591$
and needs 295. For Qwen3.5-397B-A17B these balances correspond to batches 1894
and $15\,123$ (Table~\ref{tab:balance}). These machine-balance thresholds are
not claimed service limits; they show why adding FLOPS without proportional
bandwidth worsens low-batch weight delivery.

\begin{figure}[t]
\centering
\includegraphics[width=\columnwidth]{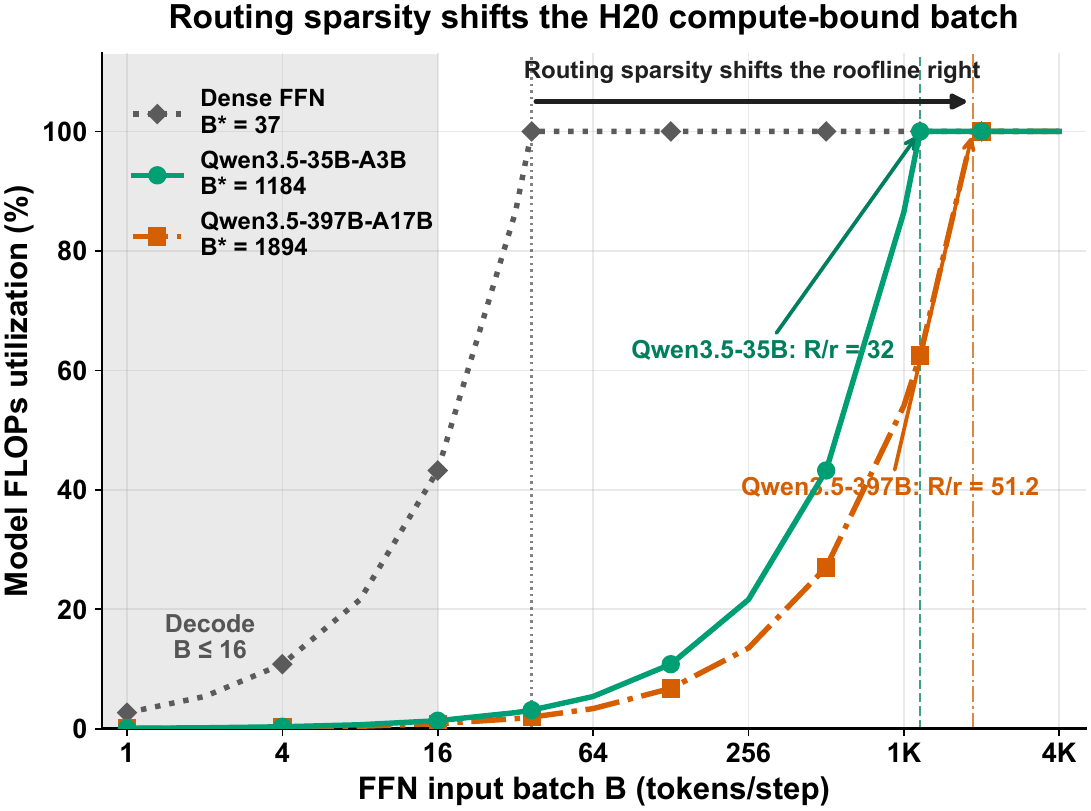}
\caption{H20 FFN MFU under dense and mean-load MoE routing.}
\label{fig:moe-roofline}
\end{figure}

Figure~\ref{fig:moe-roofline} plots H20 MFU vs FFN batch under that mean-load
model: a dense FFN reaches the roof at $B^{\star}{=}37$, while sparsity
multiplies the knee by $R/r$ (1184 for Qwen3.5-35B; ${\approx}1894$ for
397B-A17B). At $B{\leq}16$ both MoE curves stay below $1.4\%$ of peak.
These are first-order utilization knees (uniform routing, $U{\to}R$), not
measured kernel MFU; they illustrate why sparse MoE, not merely model size,
pushes commodity HBM far from the compute roof under batch-constrained decode.
H100/H800 supplies $6.7\times$ H20's dense FP8 peak at slightly less read
bandwidth, so its reuse requirement rises from 37 to 295 tokens/read and the
MoE batches become extreme (Table~\ref{tab:balance}: Qwen3.5-35B needs 1184/9452
on H20/H100; GLM-5.2 shares the $R/r{=}32$ knee at $B^{\star}{=}1184$ on H20).
The thresholds use \emph{mean} load, so under skew they are necessary
machine-balance conditions for HBM-fed execution, not service guarantees.

\begin{table}[t]
\centering
\caption{Commodity-HBM machine balance.}
\label{tab:balance}
\begin{threeparttable}
\scriptsize
\setlength{\tabcolsep}{3pt}
\begin{tabular}{@{}lrrrr@{}}
\toprule
SKU & Read BW & Peak & FLOP/B & Tok./read\tnote{a} \\
& (TB/s) & (TFLOPS) & & \\
\midrule
H20                    & 4.00 & 296  & 74  & 37 \\
H100/H800 SXM\tnote{b} & 3.35 & 1979 & 591 & 295 \\
\bottomrule
\multicolumn{5}{c}{}\\[-1.4ex]
\multicolumn{5}{@{}l}{\emph{Batch at the compute roof}} \\
\toprule
Model & $R$ & $r$ & H20 & H100/H800 \\
\midrule
Qwen3.5-35B-A3B     & 256 &  8 & 1184 & 9452 \\
Qwen3.5-397B-A17B   & 512 & 10 & 1894 & $15\,123$ \\
GLM-5.2             & 256 &  8 & 1184 & 9452 \\
\bottomrule
\end{tabular}
\begin{tablenotes}[flushleft]\footnotesize
\item[a] A read shared by $t$ tokens yields $2t$ FLOP/byte, so
$t\ge F/2$ (Eq.~\ref{eq:balance}); batch is $(F/2)(R/r)$.
\item[b] H800 shares this memory system and dense FP8 rate, differing in NVLink
bandwidth, which affects EP/TP traffic (\S\ref{sec:pacom}) rather than
per-die weight delivery.
\end{tablenotes}
\end{threeparttable}
\end{table}

\begin{figure}[t]
\centering
\includegraphics[width=\columnwidth]{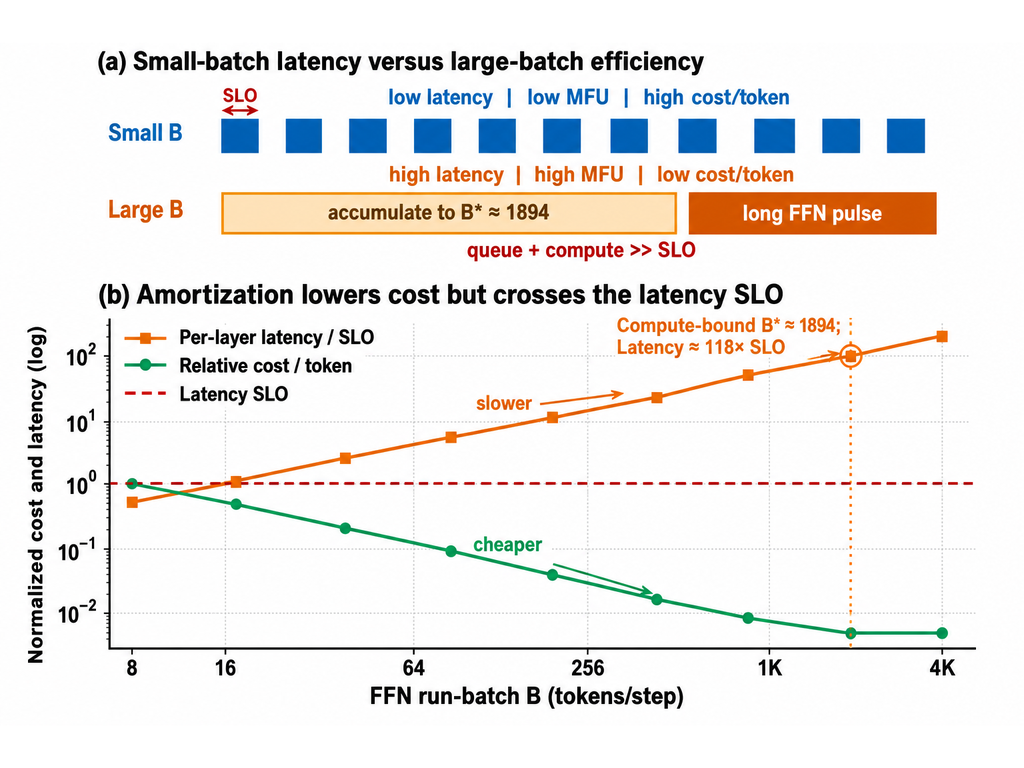}
\caption{Batch-efficiency versus latency-SLO tension for H20 MoE FFNs. With
$B{=}16$ normalized to one SLO interval, the 397B knee
($B^{\star}{\approx}1894$) sits ${\sim}118\times$ above the interactive
budget.}
\label{fig:af-batch-slo}
\end{figure}

Figure~\ref{fig:af-batch-slo} shows the serving conflict. Small batches retain
frequent low-latency pulses but leave the FFN at low MFU and high cost/token;
accumulating toward $B^{\star}$ amortizes weights and raises MFU but stretches
the gather queue into one long synchronized FFN pulse. Batch is therefore not
the primary efficiency lever for interactive decode.

\subsection{Runtime Flexibility of Near-Memory Compute}
\label{sec:bg-nmc}

Compute-in-memory (CIM) binds a stored weight tile to local arithmetic, shortening
the weight path but limiting reassignment under routing skew
\cite{shafiee2016isaac,ankit2019puma,liu2026ouroboros}. Near-memory computing
(NMC) reads weights into adjacent digital logic, paying local bandwidth to make the
mapping between weights and PE groups a runtime decision.
Heterogeneous NPU--PIM systems exploit a related split by placing bandwidth-heavy
GEMV (often attention) on DRAM-PIM and GEMM on NPUs
\cite{heo2024neupims,seo2024ianus}; hybrid DRAM/SRAM PIM further trades capacity
against latency \cite{li2026compair}; CXL near-bank PIM can remove the GPU from
the path entirely \cite{gu2025cent}.
Resident high-bandwidth NMC for \emph{MoE FFN} weights keeps experts programmed
in ReRAM while digital PEs remain reassignable under skew
(Table~\ref{tab:relatedcmp}; \S\ref{sec:design}).
CIM binds compute to programmed weights; NMC steers a local read-out to idle
digital PEs via a multicast/reduction path---the degree of freedom used for MoE
skew.

\section{Problem Analysis}
\label{sec:char}

Residency delivers weights but not useful compute: sparse routing underfills
tiles and idles PE groups.
We separate those losses on measured Qwen3.5-35B-A3B routing without extrapolating
its statistics to the larger models.
Cross-device EP/TP weight movement \cite{yu2026patterns} and expert prefetch
\cite{liu2026earth} are complementary; once weights are resident, first-order
losses are tile fill and occupancy (\S\ref{sec:pacom} bounds EP/TP traffic).

\subsection{MFU Factorization}
\label{sec:mfu}
A \emph{PE job} processes up to $w$ tokens for one expert on one PE group. For
$T$ token jobs, $J$ issued jobs, $N$ cores, $G$ groups/core, and $\rho$
critical-path rounds,
\begin{equation}
\underbrace{\frac{T}{\rho N G w}}_{\text{actual MFU}}
\;=\;
\underbrace{\frac{T}{J w}}_{\text{ideal MFU}}
\;\times\;
\underbrace{\frac{J}{\rho N G}}_{\text{occupancy}} .
\label{eq:mfu}
\end{equation}
Ideal MFU falls on underfilled tiles; occupancy falls when groups wait for a
straggler.
The two factors are reported as separate diagnostics; scheduling can affect
both.
Reporting only mean load or aggregate PE utilization would conflate the two
causes that Eq.~\ref{eq:mfu} separates.

\subsection{Routing Skew and Stability}
\label{sec:skew}\label{sec:persist}
\begin{figure}[t]
\centering
\includegraphics[width=\columnwidth]{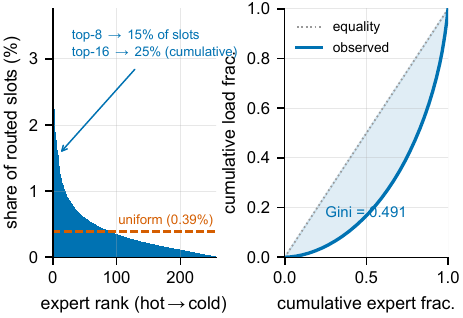}
\caption{Qwen3.5-35B-A3B at $B{=}128$: sorted expert load and Lorenz curve.
Routing is skewed (mean per-layer Gini $0.491$; top-8 slot share $15.4\%$
vs.\ $3.13\%$ uniform), creating stragglers, yet persistent enough to guide
placement (\S\ref{sec:persist}).}
\label{fig:skew_lorenz}
\end{figure}

At native $B{=}128$, top-8 routing over 256 experts has a mean expert load of
four tokens---exactly the balanced physical tile---but the observed distribution
does not place four tokens on each expert.
On the default Alpaca trace, the hottest expert receives $39.3$ tokens per step
($9.83\times$ the global mean); across the campaign the hottest/mean ratio stays
in $[6.4,9.9]\times$.
The modal hot expert persists on $40.1\%$ of decode steps
(Table~\ref{tab:skew}), enough to amortize a placement epoch.
Skew is therefore severe enough to create stragglers and stable enough to
guide placement (Fig.~\ref{fig:skew_lorenz}, Tables~\ref{tab:skew}
and~\ref{tab:campaign}; Qwen3.5-397B and GLM-5.2 serve as larger measured
models that reproduce the same skew structure and pooling knee).
Table~\ref{tab:skew} reports per-layer, per-step means over 64 decode steps;
parenthesized values are uniform-routing references.

Hot experts overflow into many full jobs while the cold tail issues 1--2-token
jobs, lowering ideal MFU even though the global token count is unchanged.
At the same time, whichever Core contains several co-fired hot experts sets the
critical-path round count, leaving other groups idle and lowering occupancy.
Persistence does not mean static routing: the signal lasts long enough to
guide a placement epoch; short-term deviations are left to queue scheduling.

\begin{table}[t]
\centering
\caption{Routing-skew statistics for the measured decode traces.}
\label{tab:skew}
\begin{threeparttable}
\setlength{\tabcolsep}{3.5pt}
\begin{tabular}{@{}lcc@{}}
\toprule
Metric (per-layer, per-step mean) & Qwen3.5-35B & Qwen3.5-397B \\
\midrule
Experts $E$ / top-$k$           & 256 / 8 & 512 / 10 \\
MoE layers                      & 40      & 60 \\
\midrule
Hottest / mean load             & $6.77\times$\tnote{a} & $12.5\times$ \\
Global peak hottest / mean      & $9.83\times$ & $47.3\times$ \\
Top-8 slot share (unif.)        & 15.4\% (3.13\%) & 12.7\% (1.56\%) \\
Top-16 slot share (unif.)       & 25.4\% (6.25\%) & 20.3\% (3.13\%) \\
Mean per-layer Gini             & 0.491 & 0.534 \\
\midrule
Hot-set persistence\tnote{b}    & 40.1\% & 49.4\% \\
\bottomrule
\end{tabular}
\begin{tablenotes}[flushleft]\footnotesize
\item[a] Default Alpaca decode traces at the native tile-fill batch
($B{=}128$ for 35B, $B{=}205$ for 397B).
Across each five-run campaign (Alpaca$\times$2 seeds, code, independent
$B{\in}\{8,64\}$), hottest/mean stays in $[6.4,9.9]\times$ (35B) and
$[12.5,18.2]\times$ (397B); Gini in $[0.47,0.58]$ and $[0.53,0.62]$.
\item[b] Fraction of decode steps whose per-layer hottest expert is the
modal hot expert of that layer.
\end{tablenotes}
\end{threeparttable}
\end{table}

\subsection{Utilization Requirements}
\label{sec:underfill}\label{sec:cb}
These observations yield four requirements.
\textbf{(i)~Preserve ideal MFU without sacrificing machine balance.}
A narrower job helps only if reduced arithmetic does not lower the roofline
faster than fill improves.
\textbf{(ii)~Recover occupancy with bounded sharing.}
A hot expert needs access to more PE groups than its owner provides, but global
sharing would move the resident weight stream across the die.
\textbf{(iii)~Control imbalance at the appropriate time scale.}
Stable coactivation guides semi-static placement; queued work fills momentary
idle groups; neither changes tile width, so gains appear in occupancy and rounds.
\textbf{(iv)~Provision communication from the mapping.}
Core-local weight multicast, Die-NoC dispatch/combining, package reduction, and
node handoff have different physical semantics; their bandwidth and reduction
order must be derived together with placement.

Increasing batch improves mean expert load but neither removes skew nor changes
which groups can consume a resident read-out, and service latency bounds this
form of amortization.
\S\ref{sec:design} implements these requirements;
\S\ref{sec:eval} reports the measured pooling knee and mechanism gains.

\section{Architecture}
\label{sec:design}
\subsection{Architecture Overview and Design Contract}
\begin{figure*}[t]
\centering
\includegraphics[width=\textwidth]{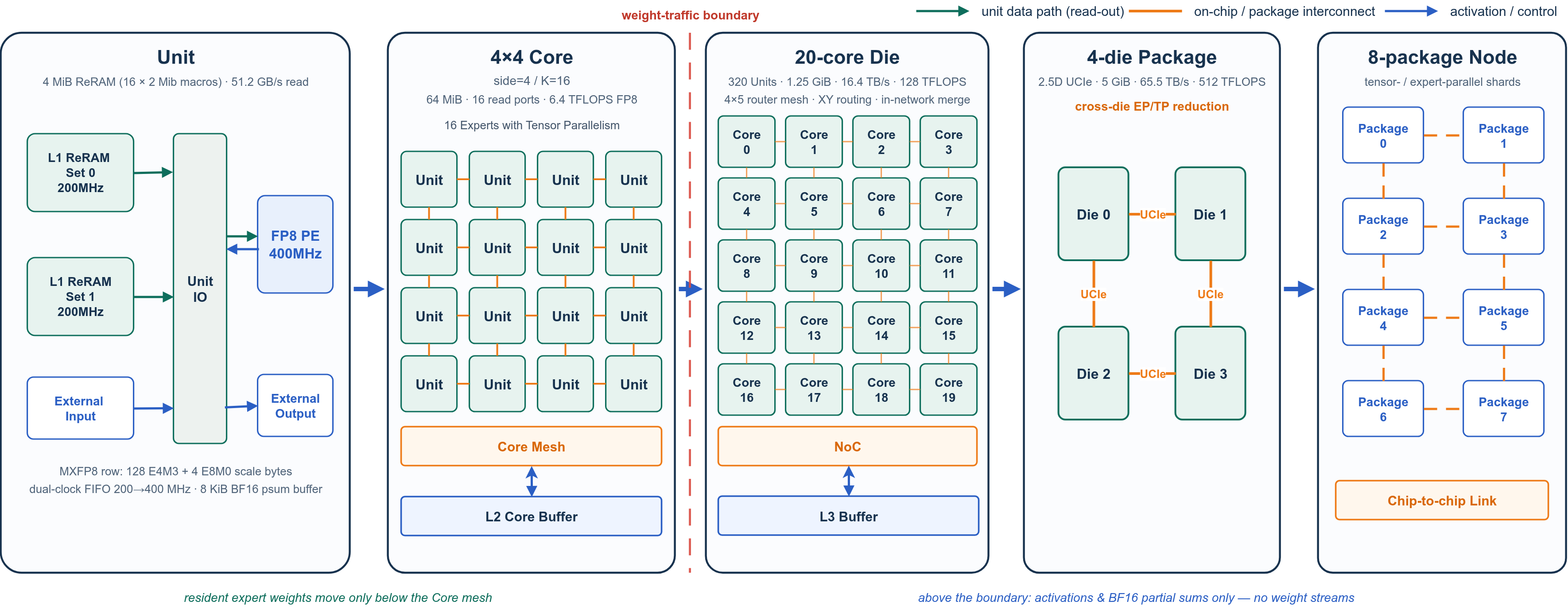}
\caption{ReXpert hierarchy (Unit $\to$ Core $\to$ Die $\to$ Package $\to$
Node): expert weights stay at/below the Core mesh; higher levels carry only
activations and partials.}
\label{fig:hw_architecture}
\end{figure*}

ReXpert is a resident ReRAM near-memory FFN pool organized as
Unit$\rightarrow$Core$\rightarrow$Die$\rightarrow$4-die 2.5D UCIe Package$\rightarrow$Node
(Fig.~\ref{fig:hw_architecture}).
Each Node connects eight such Packages.
Its contract follows the MFU cascade of Eq.~\ref{eq:mfu}.
The Unit IO and PE expose a balanced four-token tile that sets the roofline
within which \emph{ideal MFU} measures issued-tile fill.
The core mesh and scheduler reclaim idle PE-group time, raising
\emph{occupancy}; their product is \emph{actual MFU}.
Thus tile width addresses the ideal-MFU ceiling, whereas bounded pooling,
placement, and token fetch address occupancy.

The interface boundary is deliberately local: Unit IO may redirect a resident
expert's read-out across its core, but weights do not traverse links above the
core mesh.
This keeps the bandwidth-density advantage of residency while allowing idle PE
groups to serve a hot expert.
Larger sharing domains recover more occupancy but require a wider, longer
multicast/reduction fabric; ReXpert therefore bounds sharing within a core and
uses placement and queue scheduling beyond that boundary.

A Unit is the smallest independently readable weight store and compute endpoint.
A Core groups 16 such endpoints so a resident read-out can be reused by PE
groups that do not own the weight.
The Die NoC connects cores for activation dispatch and partial-sum collection;
package and node links connect tensor- and expert-parallel shards.
Widening the core-local domain lengthens every multicast tree; traffic above
the core carries compact vectors, so the core boundary is both a scheduling
boundary and the only one across which resident weights are observed.

\subsection{Balanced ReXpert Unit}
\label{sec:design-unit}
A ReXpert unit stores $4$\,MiB in 16 $2$\,Mib macros
($16\mathrm{K}\times128$\,b).
The macros operate at $200$\,MHz, providing $16\times128$\,b per cycle and
$51.2$\,GB/s of local read bandwidth.
Eight even/odd address groups assemble a $2048$\,b frame.
Unit IO crosses that frame through a dual-clock FIFO into the $400$\,MHz PE
domain and repacks it into MXFP8 rows: 128 E4M3 elements plus four E8M0 scale
bytes.
The same interface selects a local PE destination or injects the read-out into
the core mesh.

The PE is a four-token by 128-lane FP8 array.
It performs 512 MACs per cycle, or approximately $400$\,GFLOPS, while holding
four activation vectors stationary and broadcasting each weight row across them.
Fixed-point accumulators emit BF16 results through an $8$\,KiB partial-sum
buffer, which is also the unit's core-mesh boundary.

\paragraph{Dataflow and balance.}
For a four-token job, Unit IO installs four activation rows in the PE and
advances the resident expert shard one 128-element weight row at a time.
Each row is consumed by all four token lanes in the 400\,MHz domain while the
200\,MHz macros prepare the next frame.
Accumulation remains local across the reduction dimension; only the completed
BF16 partial vector enters the output buffer.
Gate and up-projection shards follow this path, the digital PE applies the
activation, and the down-projection shard produces the vector reduced at the
next hierarchy level.
Ordinary execution therefore spends neither Die-NoC nor package bandwidth on
weights.
Pooling changes only the destination of a frame: Unit IO may inject the same
read-out into the Core mesh so borrowed PE groups apply it to different
four-token jobs.

The unit balance is $400/51.2=7.81$\,FLOP/byte.
A width-$w$ input-stationary tile offers $2w$\,FLOP/byte, so Eq.~\ref{eq:balance}
requires $2w\ge7.81$; $w{=}4$ is the narrowest integer width that can reach peak
compute.
Every hierarchy level scales compute and local read bandwidth by the same unit
count, preserving $0.128$\,bytes/FLOP.

\paragraph{Hierarchy instantiation.}
The reference model uses a $20{\times}20$\,mm 12\,nm die budgeted as
$268$\,mm$^2$ ReRAM, $50$\,mm$^2$ NoC/D2D, and $82$\,mm$^2$ logic, using
reported same-node 1T1R macro density \cite{tsmc2022reram12nm}.
A die contains 320 units as 20 side-4 cores ($1.25$\,GiB, $16.4$\,TB/s,
$128$\,TFLOPS FP8).
Four dies form a $5$\,GiB, $65.5$\,TB/s, $512$\,TFLOPS package; the reference
FFN pool uses 25 dies ($31.25$\,GiB, $410$\,TB/s, $3.2$\,PFLOPS;
Table~\ref{tab:hwconfig}), sized to hold the $30.1$\,GiB of Qwen3.5-35B FFN
weights.
Following IANUS-class practice of closing area/power as a first-class result
\cite{seo2024ianus}, Table~\ref{tab:overhead} (\S\ref{sec:eval-cost}) gives
the die area and power budget.
The 320 Units form a $16{\times}20$ array, so a side-4 Core tiles the die
without remainder and exposes 20 Core routers as a $4{\times}5$ Die-NoC mesh.
\begin{table}[t]
\centering
\caption{ReXpert hardware hierarchy.}
\label{tab:hwconfig}
\begin{threeparttable}
\scriptsize
\setlength{\tabcolsep}{2.5pt}
\begin{tabular}{@{}p{0.13\columnwidth}p{0.28\columnwidth}p{0.20\columnwidth}p{0.16\columnwidth}p{0.11\columnwidth}@{}}
\toprule
Level & Resident capacity & Local read BW & Peak FP8 & B/FLOP \\
\midrule
Unit &
$4$ MiB &
$51.2$ GB/s &
$400$ GFLOPS & $0.128$ \\
Core &
$4{\times}4$ units, $64$ MiB &
$819.2$ GB/s &
$6.4$ TFLOPS & $0.128$ \\
Die &
$16{\times}20$ units, $20$ cores, $1.25$ GiB &
$16.4$ TB/s &
$128$ TFLOPS & $0.128$ \\
Package &
$4$ dies, $5$ GiB &
$65.5$ TB/s &
$512$ TFLOPS & $0.128$ \\
Node &
$8$ chips, $40$ GiB &
$524$ TB/s &
$4.1$ PFLOPS & $0.128$ \\
\bottomrule
\end{tabular}
\begin{tablenotes}[flushleft]\footnotesize
\item Read bandwidth sums local ReRAM ports and excludes interconnect (Table~\ref{tab:commbudget}).
Unit primitives:
$16\times128$\,b\,@\,$200$\,MHz $=51.2$\,GB/s and a $4$-token $\times$
$128$-lane FP8 PE\,@\,$400$\,MHz $=400$\,GFLOPS; upper levels scale by unit
count, holding B/FLOP constant (\S\ref{sec:design-unit}).
\end{tablenotes}
\end{threeparttable}
\end{table}

\subsection{Core-Local Pooling and Broadcast}
\label{sec:core}
\begin{figure*}[t]
\centering
\includegraphics[width=\textwidth]{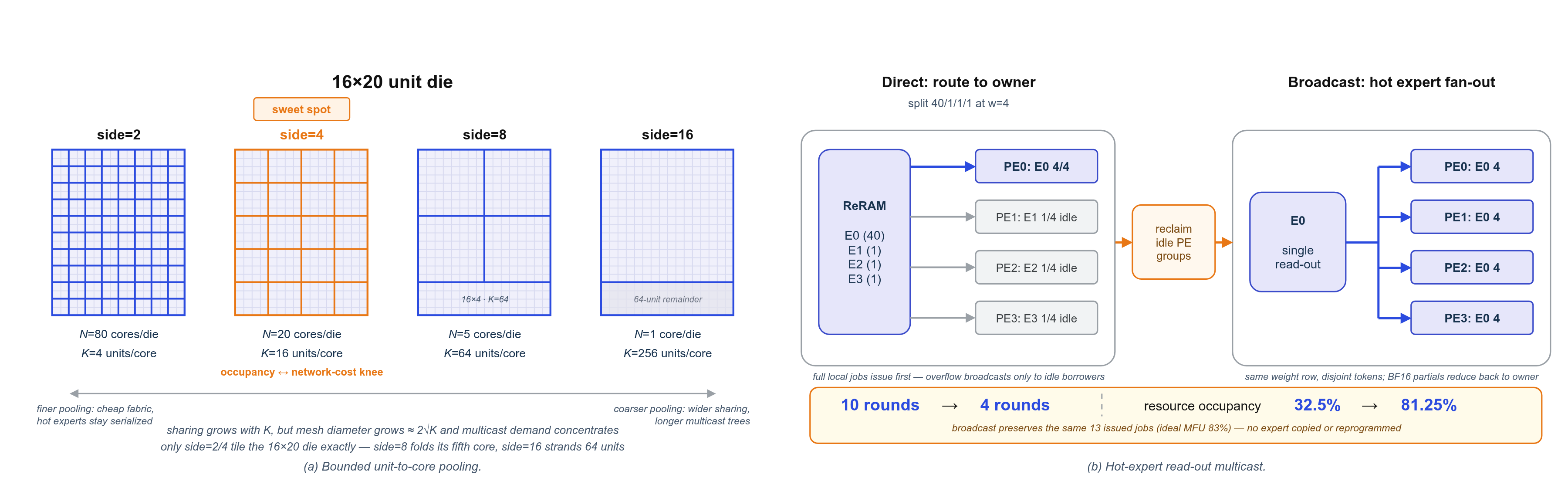}
\caption{Core-local pooling. (a) Sharing is bounded to the core: idle PE
groups pool for one resident hot expert. (b) Its read-out is multicast to
borrowers holding disjoint tokens; no weight is replicated---the gain is
fewer critical-path rounds.}
\label{fig:pooling_dataflow}
\end{figure*}
A core connects $K=\mathrm{side}^2$ units through a multicast/reduction mesh.
Its $K$ PE groups may execute their local experts independently, or borrow
groups for one resident hot expert.
The selected side-4 core therefore combines 16 units, $64$\,MiB, 16 read ports,
and 16 PE groups.
Pooling changes neither the physical tile nor the jobs it issues; it changes how
many groups can make progress on those jobs concurrently.

\paragraph{One routed wave.}
The placement table maps each routed $(\text{token},\text{expert})$ pair to the
Core that stores the corresponding shard.
The Die NoC delivers the activation to that Core; the Core scheduler groups pairs
by expert into width-four jobs and records the owner Unit for each resident
read-out.
Full local jobs begin immediately; underfilled jobs remain valid; overflow from
a hot owner enters broadcast only after the scheduler identifies idle
borrowers.
This ordering keeps the common path point-to-point and invokes the high-rate mesh
only for imbalance.

\paragraph{Two-stage broadcast dataflow.}
\label{sec:dataflow}
The scheduler first serves resident experts through each owner's local
input-stationary path.
If one expert has more tiles than its neighbors, a second stage multicasts that
expert's read-out and assigns remaining tokens to idle groups.
As a worked example matching the measured skew (\S\ref{sec:char}),
consider one hot expert with $40$ queued jobs co-resident with three cold
experts holding one job each at $w{=}4$: direct execution takes ten
rounds: the three one-token experts each issue one underfilled local job, while
the hot expert serializes ten full jobs, giving occupancy
$13/(10{\times}4)=32.5\%$.
Broadcast preserves the same 13 issued jobs (ideal MFU $83\%$) but schedules them
over four groups in four rounds at $81.25\%$ occupancy.
Borrowers receive the same weight row but hold disjoint token activations; their
BF16 partials reduce back to the owner before one result leaves the Core.
No expert is copied or reprogrammed---the gain is shorter critical-path rounds
and fewer repeated local reads of the same weights
(Fig.~\ref{fig:pooling_dataflow}).

\paragraph{Bounded pooling degree.}
\label{sec:knee}
Increasing $K$ exposes more groups to a hot read-out, but mesh diameter grows
approximately as $2\sqrt{K}$ and each larger core concentrates multicast demand.
ReXpert fixes side\,$=$\,4 as the bounded sharing domain: it tiles the
$16\times20$ unit floorplan into 20 cores while keeping weight movement local.
\S\ref{sec:pacom} sizes the corresponding multicast fabric, and
\S\ref{sec:eval-balance} evaluates the occupancy--network-cost knee.
On the default Qwen3.5-35B-A3B Alpaca trace, side-4 pooling raises occupancy from
$0.328$ at side~2 to $0.519$ and actual MFU from $0.240$ to $0.381$ (ideal MFU
stays at $0.736$---mean load is four tokens/expert, but the cold tail underfills
tiles); larger domains suffer $1.69$--$2.50\times$ round stretch.
After network cost, side~4 minimizes silicon-per-delivered-occupancy.

\subsection{Tile Width and Cross-Core Balance}
\label{sec:tilewidth}\label{sec:mapping}\label{sec:fetch}
Narrowing $w$ appears to improve ideal MFU because fewer token slots must be
filled, but it simultaneously lowers arithmetic intensity.
On the balanced unit, the attainable peak fraction is capped by
$\min(1,2w/7.8125)$: $0.256$ for $w{=}1$, $0.512$ for $w{=}2$, and one for
$w{=}4$.
The relevant quantity is therefore the \emph{roofline-adjusted ideal MFU};
narrow masks also require more read-outs for the same token set.
Hence $w{=}4$ remains the balanced physical tile, while pooling recovers
occupancy without sacrificing its reuse.
The measured width sweep is reported in \S\ref{sec:eval-balance}.

Across cores, pooling absorbs skew within a core; placement controls which
experts share that finite budget.
ReXpert estimates expert load and pairwise coactivation from a profiling window,
spreads frequently co-fired hot experts across cores, and pairs hot experts with
colder ones within each core.
Placement is semi-static: it changes metadata rather than rewriting resident
ReRAM weights.
At the selected side-4, $G{=}16$ point, this policy raises measured occupancy
$0.519\to0.532$ and reduces rounds from $2.16$ to $2.07$
relative to load-only LPT placement.

Load-aware token fetch is the dynamic complement.
When a core would otherwise be idle, the scheduler may fetch queued tokens
already routed to its resident experts and execute them ahead of FIFO order.
Unlike EARTH's speculative expert prefetch, which changes what weights move
\cite{liu2026earth}, fetch here only reorders tokens already routed to resident
experts inside a bounded continuous-batch queue.
With only $G{=}2$ PE groups per core, a depth-$16\times$ load-aware queue
raises occupancy $0.422\to0.552$; at side-4 with $G{=}16$, pooling has
already absorbed most of that imbalance.
Fetch is therefore a bounded latency--occupancy exchange for residual queue
variation, not an additional source of peak throughput.

\paragraph{Summary.}
The balanced $w{=}4$ tile protects ideal MFU; occupancy is recovered primarily
by bounded multicast, with placement and fetch providing secondary recovery;
weights appear on the Core mesh only as a bounded local exception.
\S\ref{sec:pacom} sizes the fabrics these mechanisms induce.

\section{Parallelism and Interconnect Analysis}
\label{sec:pacom}

\begin{figure*}[t]
\centering
\includegraphics[width=\textwidth]{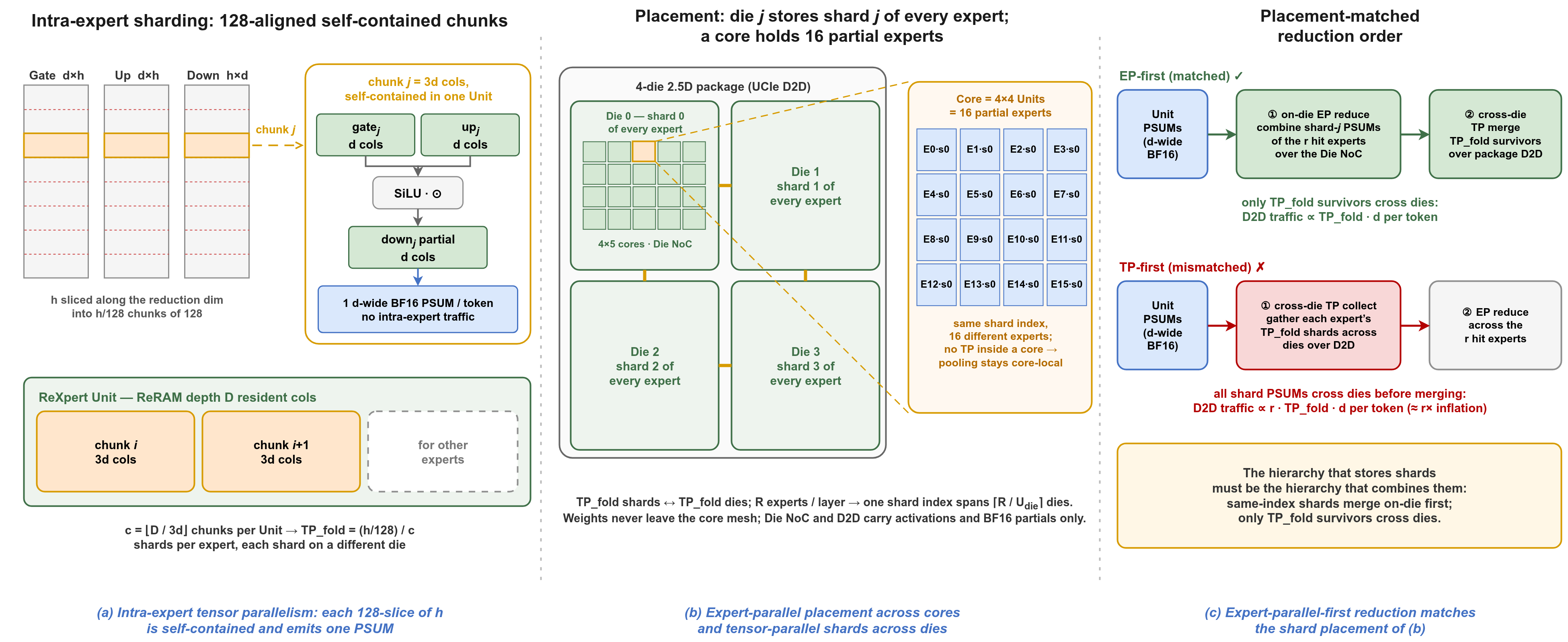}
\caption{Parallel mapping and reduction order.
(a)~Intra-expert tensor parallelism: each 128-slice of the reduction dimension
is a self-contained chunk that emits one $d$-wide BF16 PSUM.
(b)~Placement: die $j$ stores shard $j$ of every expert and a core holds 16
partial experts, keeping pooling core-local.
(c)~Expert-parallel-first reduction matches this placement; the reversed order
inflates package D2D traffic up to $5.7\times$.}
\label{fig:pacom_mapping}
\end{figure*}

Residency removes sustained weight traffic above the core mesh; pooling is the
local exception that multicasts hot-expert read-outs inside a core.
We size each level from demand that can overlap the resident array stream:
\begin{align}
 t_{\mathrm{layer}}&=D_{\mathrm{used}}t_{\mathrm{col}},\quad
 B_\ell^{\mathrm{req}}=V_\ell/t_{\mathrm{layer}}, \nonumber\\
 V_\ell&=n_{\mathrm{tok}}\bigl(\phi_{\ell}^{\mathrm{act}}d
          +\phi_{\ell}^{\mathrm{psum}}2d\bigr),
\end{align}
with $t_{\mathrm{col}}=2.5$\,ns and $\phi$ counting dispatch/reduction after
in-network combining.
Experts are chunked along the reduction dimension into $128$-aligned shards; die
$j$ stores shard $j$ of every expert and a core holds 16 partial experts, which
fixes the matched reduction order below (Fig.~\ref{fig:pacom_mapping}).
Average demand does not bound bursts, so the Core mesh is checked at p99 and peak.

\subsection{Per-Level Semantics}
\label{sec:pacom-levels}
Each level has a distinct physical job; provisioning one aggregate bandwidth
number cannot establish that pooling remains local.

\textbf{Unit ports.}
A four-token wave exchanges $4d$/$8d$ activation/partial bytes on a private path
that never carries other Units' weights.

\textbf{Core mesh.}
This is the only weight-carrying fabric: cold experts stay local; hot read-outs
multicast and BF16 partials reduce to one Core endpoint before leaving the Core.

\textbf{Die NoC.}
A $4{\times}5$ mesh of 20 cores carries activation dispatch and combined
partials; busiest-link demand is a routed quantity after Core-local combining.

\textbf{Package D2D and node.}
These links move activations/partials and the inter-layer residual only---never
steady-state FFN weights.
CHIME provisions the complementary attention pool \cite{liu2026chime}; multi-unit
MoE EP movement is orthogonal once weights are resident \cite{yu2026patterns}.

\subsection{Provisioning Results}
\label{sec:pacom-results}\label{sec:pacom-node}\label{sec:pacom-dse}
\begin{table}[t]
\centering
\caption{Per-level communication demand and selected provisioning.}
\label{tab:commbudget}
\begin{threeparttable}
\scriptsize
\setlength{\tabcolsep}{2.5pt}
\begin{tabular}{@{}p{0.23\columnwidth}p{0.27\columnwidth}
                p{0.21\columnwidth}p{0.19\columnwidth}@{}}
\toprule
Path & Demand & Selected provision & Evidence scope \\
\midrule
Unit external I/O &
$1.57$ GB/s &
$51.2$ GB/s &
modeled \\
Core mesh &
$51.2$ GB/s/stream;
$204.8$ p99 &
$64$ GB/s/link;
$256$ GB/s/core &
Qwen3.5 trace \\
Die NoC, busiest link &
$24.68$--$24.81$ GB/s\tnote{b} &
$64$ GB/s/link &
GLM BookSim2 \\
Package D2D, per die &
$47.0$ GB/s &
$96$ GB/s &
modeled \\
Node residual &
$16.7$ GB/s &
$34$ GB/s &
modeled \\
\bottomrule
\end{tabular}
\begin{tablenotes}[flushleft]\footnotesize
\item[a] Qwen3.5-35B-A3B aggregate core-mesh demand.  The selected
$256$ GB/s covers p99.
\item[b] The two values are expert-parallel-first and
tensor-parallel-first for GLM-5.2.  Both on-die links are $512$ b at 1 GHz.
\end{tablenotes}
\end{threeparttable}
\end{table}

BookSim2 shows GLM-5.2's Die-NoC busiest link at $24.7$--$24.8$\,GB/s, so a
$512$\,b/1\,GHz ($64$\,GB/s) link is the narrowest viable point.
The Core mesh is sized by an indivisible $51.2$\,GB/s read-out ($\ge$$410$\,b at
1\,GHz; we select $512$\,b).
Qwen3.5-35B concurrent multicast reaches $204.8$\,GB/s p99 and $307.2$\,GB/s
peak; the
selected $256$\,GB/s per-core budget covers p99 (five streams), not the peak.
Unicast replication is infeasible ($15\times$ source port; $3.6\times$ traffic),
so multicast is a pooling prerequisite.
At this fabric point the joint DSE selects $4{\times}4$ cores ($44\%$ less
silicon per delivered throughput than the next-best size).
DSENT with 12\,nm-scaled device parameters puts the Core-mesh area at
$35.6$\,mm$^2$, within the $50$\,mm$^2$ NoC/D2D budget.
Off-die modeled demand is $47.0$\,GB/s D2D/die and $16.7$\,GB/s node residual,
provisioned at $96$ and $34$\,GB/s (Table~\ref{tab:commbudget}).

\subsection{Placement-Matched Reduction Order}
\label{sec:pacom-order}\label{sec:pacom-limits}
Expert-parallel-first reduction merges same-index shards on-die before the
tensor-parallel merge crosses dies.
The reverse order inflates GLM-5.2 intra-package D2D from $37.6$ to
$213.1$\,GB/s ($5.7\times$) while barely moving Die-NoC load
($24.68\to24.81$\,GB/s); Qwen3.5-397B and Qwen3.5-35B see $2.7\times$ and
$5.7\times$ D2D penalties.
Favorable provisioning therefore requires matching reduction order to placement:
the hierarchy that stores shards must also be the hierarchy that combines them.

\subsection{Array-Depth Design Space}
\label{sec:pacom-depth}
Array depth sets storage density at a fixed column rate, and therefore
the hide time available to communication.
The Unit in \S\ref{sec:design-unit} stores $4$\,MiB as $32$K weight
columns and streams them at $t_{\mathrm{col}}{=}2.5$\,ns
($51.2$\,GB/s).
A shallower array finishes the resident read sooner, emits the BF16
partial sooner, and shortens $t_{\mathrm{layer}}$, so
$B_\ell^{\mathrm{req}}{=}V_\ell/t_{\mathrm{layer}}$ rises.
A $128$-aligned chunk costs $3d$ columns; below that quantum the expert
folds into more tensor-parallel shards and cross-die volume grows as
well.

\begin{figure}[t]
\centering
\includegraphics[width=\columnwidth]{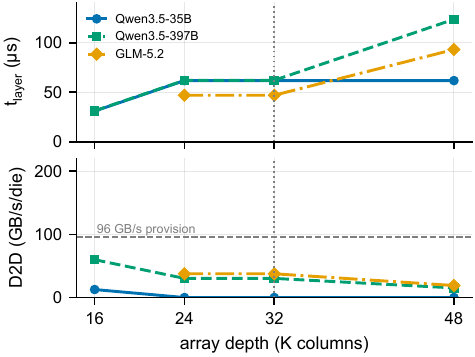}
\caption{Array depth at fixed column rate. Top: $t_{\mathrm{layer}}$
shortens at 16K, holds while $D_{\mathrm{used}}$ is unchanged
(24K--32K), then stretches when TP coarsens (48K). Bottom: D2D demand
is the inverse; the dashed line is the $96$\,GB/s/die provision.}
\label{fig:depth_dse}
\end{figure}

Fig.~\ref{fig:depth_dse} sweeps depth at that fixed rate.
On Qwen3.5-35B, $16$K versus $32$K halves layer time
($61.8{\to}31.1\,\mu$s), opens a D2D path the single-shard mapping did
not need ($0{\to}12.7$\,GB/s/die), doubles the busiest Die-NoC link, and
grows the pool $25{\to}49$ dies.
Qwen3.5-397B shows the same $2.0\times$ time--bandwidth exchange
(D2D $30.3{\to}59.9$\,GB/s/die; TP $4{\to}8$).
GLM-5.2 cannot map at $16$K: one chunk is $18432$ columns.
Deepening to $48$K reverses the trade---$397$B and GLM stretch the pulse
by $2.0\times$ and cut D2D roughly in half, reducing die count
($289{\to}193$, $543{\to}362$).
$24$K is a packing point, not a latency point: $35$B and $397$B reach
$100\%$ array utilization at the same $t_{\mathrm{layer}}$ as $32$K
because occupied columns do not grow.

The $32$K Unit is a feasible density that keeps modeled D2D under the
$96$\,GB/s provision of Table~\ref{tab:commbudget} for all three models.
A shallower array is faster only if the hierarchy is resized with it;
\S\ref{sec:pacom-results} reports the $32$K budget.

\section{Evaluation Methodology}
\label{sec:method}

All quantitative claims in this paper are a
\emph{measured$+$modeled hybrid analysis}:
router-hook MoE traces and GPU grouped-GEMM kernel calibration supply the
measured inputs; a kernel-calibrated roofline, BookSim2$+$DSENT networks, and
an RTL/DSENT/UCIe die cost model project the ReXpert FFN pool and AF system.
We do not claim silicon measurements of ReXpert itself.

\paragraph{Workloads and measurement scope.}
We evaluate three FP8 MoE models spanning the deployment range:
Qwen3.5-35B-A3B (256 experts, top-8, 40 layers),
Qwen3.5-397B-A17B (512 experts, top-10, 60 layers), and
GLM-5.2 (256 experts, top-8, 75 MoE layers).
Skew, pooling, placement, fetch, and expert-union inputs use router-hook
measurements on all three models: Qwen3.5-35B on eight H20 GPUs, and
Qwen3.5-397B and GLM-5.2 on their serving clusters.
To cover workload diversity and batch effects
\cite{liu2026chime,yu2026patterns,heo2024neupims}, each campaign captures
Alpaca at the native tile-fill batch with two seeds, a HumanEval/code batch
at the same batch, and \emph{independent} Alpaca runs at
$B{\in}\{8,64\}$ (Table~\ref{tab:campaign} details 35B and 397B; GLM-5.2
uses the same protocol at $B{=}205$).
The default trace is Alpaca decode at the native tile-fill batch
($B{=}128$ for 35B; $B{=}205$ for 397B and GLM-5.2), which sets the mean
tokens/expert to four---matching the $w{=}4$ physical tile.
Unless a table states otherwise, mechanism numbers (ideal MFU, occupancy,
straggler, fetch) are from these traces; capacity and interconnect sizing
use the corresponding model shapes.
We do not re-implement Tutel/FasterMoE/MoE-Lightning as quantitative
baselines: on decode their throughput is bounded above by the HBM roofline we
already grant the GPU hardware; Table~\ref{tab:relatedcmp} positions them
qualitatively.

\paragraph{Performance model.}
Per-expert latency is
$\max(t_{\mathrm{compute}},t_{\mathrm{weight\ read}})+t_{\mathrm{comm}}$
\cite{williams2009roofline,cao2025moelightning}, using
Tables~\ref{tab:hwconfig} and \ref{tab:commbudget}.
Weight traffic is $L(U(B)P_e+\text{router})$ bytes per step; measured per-layer
unions match the independence model within $3\%$ on the default trace.
BookSim2+DSENT models the Core mesh and Die NoC; pool latency, energy, and
AF TPOT are therefore projected from measured traces and calibrated
parameters, not from a ReXpert prototype.
The array-depth sweep (\S\ref{sec:pacom-depth}) holds the Unit column
rate fixed at $51.2$\,GB/s and varies only resident columns
($16$K--$48$K), so $t_{\mathrm{layer}}$ and
$B_\ell^{\mathrm{req}}$ move together; mapping A and the EP-first
reduction of \S\ref{sec:pacom} are reused.
Both sides use the same straggler-free roofline, and the GPU side is
calibrated: grouped-GEMM decode kernels replaying the captured routing reach
$88$--$94\%$ of datasheet HBM stream bandwidth and the roofline
matches measured kernel time within $7\%$.
Baselines are credited with \emph{full} stream bandwidth and no kernel-launch
or software overhead throughout, so cross-system gaps are architecture-level
bandwidth-density bounds (\S\ref{sec:eval-mfu} reports scheduling from the
same measured traces).
The AF-disaggregated system (\S\ref{sec:eval-afslo}) pairs the same FFN
roofline with an H20 attention stage (proj$+$KV) under iso-compute die counts,
in the spirit of CHIME's A/F sizing: the AF pipeline overlaps stages
(TPOT $=\max(t_{\mathrm{attn}},t_{\mathrm{FFN}})$), while the homogeneous-H20
baseline runs both stages on the same dies
(TPOT $=t_{\mathrm{attn}}+t_{\mathrm{FFN}}$).
Attention shapes for all three models come from the served checkpoint
configurations.
ReRAM bandwidth sensitivity scales the nominal ReXpert read BW at fixed
peak compute.

\paragraph{Baselines and normalization.}
Table~\ref{tab:baselines} applies the same roofline to H20 and H100/H800 pools
at equal capacity, equal peak compute, and equal read bandwidth
(Table~\ref{tab:norm}); equal peak compute is the headline.
Table~\ref{tab:efficiency} isolates weight-movement energy; system power and
energy per token are modeled end-to-end over the FFN pool
(\S\ref{sec:eval-cost}).

\paragraph{Metrics.}
Eq.~\ref{eq:mfu} defines ideal MFU, occupancy, and actual MFU.
Mesh studies also report round stretch and
effective MFU $=\text{actual MFU}/\text{stretch}$.
Tile-width uses \emph{roofline-adjusted ideal MFU} (fill $\times$ width cap).
Default: $N{=}20$ cores/die, $G{=}16$ PE groups/core, tile width $w{=}4$,
broadcast at the selected $4{\times}4$
point.
\begin{table}[t]
\centering
\caption{MoE FFN-pool hardware baselines.}
\label{tab:baselines}
\begin{threeparttable}
\scriptsize
\setlength{\tabcolsep}{3pt}
\begin{tabular}{@{}lcccccc@{}}
\toprule
& Cap. & Read BW & Peak & \multicolumn{3}{c}{Dies to equalize\tnote{a}} \\
\cmidrule(lr){5-7}
System & /die & /die & /die & Cap. & Comp. & BW \\
       & (GiB) & (TB/s) & (TF) & & & \\
\midrule
H100/H800 SXM\tnote{b} & 80 & 3.35 & 1979 & 1 & 1.6 & 122 \\
H20\tnote{b}           & 96 & 4.00 & 296  & 1 & 10.8 & 102  \\
\textbf{ReXpert} & \textbf{1.25} & \textbf{16.4} & \textbf{128}
  & \textbf{25} & \textbf{25} & \textbf{25} \\
\bottomrule
\end{tabular}
\begin{tablenotes}[flushleft]\footnotesize
\item[a] Dies to match our 25-die pool, sized for the $30.1$\,GiB FFN weights,
on capacity, peak compute ($3200$ TFLOPS), or read bandwidth ($410$\,TB/s).
We report all three
because no single one is neutral; Table~\ref{tab:norm} gives the latencies each
implies.
\item[b] Dense FP8, not the sparsity-inflated datasheet peaks. H800 shares the
H100 SXM memory system and FP8 rate, differing in NVLink ($400$ vs $900$\,GB/s).
\end{tablenotes}
\end{threeparttable}
\end{table}

\begin{table}[t]
\centering
\caption{Routing-campaign robustness (64 decode steps per run).}
\label{tab:campaign}
\begin{threeparttable}
\scriptsize
\setlength{\tabcolsep}{2.6pt}
\begin{tabular}{@{}lllcccc@{}}
\toprule
Model & Workload & $B$ & Seed & Gini & Hot/mean & $U(B)$ \\
\midrule
\multirow{5}{*}{35B}
 & Alpaca (default) & 128 & 0 & 0.491 & $6.77\times$ & 205.0 \\
 & Alpaca           & 128 & 1 & 0.471 & $6.58\times$ & 210.2 \\
 & Code (HumanEval) & 128 & 0 & 0.522 & $9.89\times$ & 202.5 \\
 & Alpaca (indep.)  & 8   & 0 & 0.583 & $9.31\times$ & 50.1 \\
 & Alpaca (indep.)  & 64  & 0 & 0.480 & $6.43\times$ & 173.1 \\
\midrule
\multirow{5}{*}{397B}
 & Alpaca (default) & 205 & 0 & 0.534 & $12.5\times$ & 376.2 \\
 & Alpaca           & 205 & 1 & 0.536 & $13.3\times$ & 377.6 \\
 & Code (HumanEval) & 205 & 0 & 0.554 & $18.2\times$ & 370.3 \\
 & Alpaca (indep.)  & 8   & 0 & 0.619 & $13.7\times$ & 65.1 \\
 & Alpaca (indep.)  & 64  & 0 & 0.558 & $13.1\times$ & 250.7 \\
\bottomrule
\end{tabular}
\begin{tablenotes}[flushleft]\footnotesize
\item Default batch is the native tile-fill point ($B^{*}{=}4E/k$: 128 for
35B, 205 for 397B).
The $B{\in}\{8,64\}$ captures are independent serving runs, not derived
from the default trace; sliced $U(B)$ from the default trace matches these
natives within $4\%$ on 35B ($47.9$ vs $50.1$ at $B{=}8$) and $1\%$ on 397B
($65.4$ vs $65.1$ at $B{=}8$).
Code is more skewed than Alpaca at the matched default batch, but $U(B)$
stays within $4\%$ of the Alpaca union on both models.
\end{tablenotes}
\end{threeparttable}
\end{table}

\section{Evaluation}
\label{sec:eval}
We report a measured$+$modeled hybrid analysis
(\S\ref{sec:method}): measured traces drive MFU-cascade mechanisms; calibrated
models project bandwidth-density gains, interconnect headroom, AF serving, and
area/power across the three-model range.

\subsection{RQ1: FFN-Pool Latency and Weight-Movement Energy}
\label{sec:eval-cap}
\begin{figure}[t]
\centering
\includegraphics[width=\columnwidth]{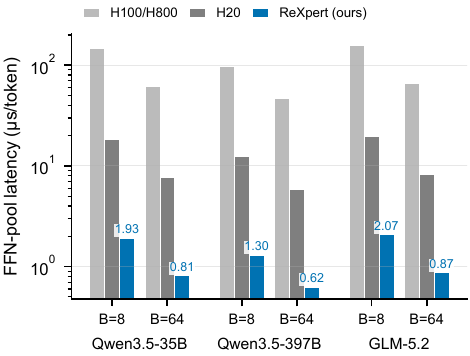}
\caption{Iso-peak-compute FFN-pool latency per token across the three models. With both pools read-bound, the gaps track
the read-bandwidth-density ratios---$9.5\times$ vs H20, $75.6\times$ vs
H100/H800.}
\label{fig:e2e_multimodel}
\end{figure}

At iso-peak-compute (Fig.~\ref{fig:e2e_multimodel}), ReXpert reduces
FFN-pool latency by $9.5\times$ vs H20 and $75.6\times$ vs
H100/H800 on Qwen3.5-35B at $B{=}8$ ($1.93$ vs $18.26$/$145.8\,\mu$s/token).
The same resident-pool construction scales to
Qwen3.5-397B-A17B (289 dies, $1.30\,\mu$s/token) and
GLM-5.2 (543 dies, $2.07\,\mu$s/token) on their measured traces: because every hierarchy level holds
$0.128$\,B/FLOP, both sides stay read-bound through $B{=}64$ on all three
models and the iso-compute ratio remains $9.5\times$/$75.6\times$
independent of model scale.
This constancy follows from both sides being read-bound: the iso-compute gap
equals the read-bandwidth-density ratio of the two memory systems, a property
of weight residency rather than of scheduling.
These are FFN-pool results, not whole-model TPOT
\cite{gu2025cent,liu2026chime}; \S\ref{sec:eval-afslo} presents the
system-level comparison.

Both systems move $331$\,MB/token of activated weights at $B{=}64$; at
$4.00$ vs $0.20$\,pJ/byte the store-energy ratio is exactly $20\times$
(Table~\ref{tab:efficiency}).
Under equal-bandwidth normalization (Table~\ref{tab:norm}), matching our
aggregate read bandwidth requires ${\sim}100$ GPU dies and erases the latency
advantage: the gain arises from bandwidth \emph{density}, not aggregate
bandwidth.
Scaling ReRAM BW by $0.5$--$2\times$ at fixed peak compute moves the
$B{=}8$ H20 gap from $4.7\times$ to $18.9\times$, confirming that the
advantage tracks read bandwidth while ReXpert stays read-bound.

\begin{table}[t]
\centering
\caption{Capacity and weight-movement energy at $B{=}64$.}
\label{tab:efficiency}
\begin{threeparttable}
\scriptsize
\setlength{\tabcolsep}{3.5pt}
\begin{tabular}{@{}lccccc@{}}
\toprule
       & Dies\tnote{a} & Weight & Off-chip & Weight-move & Rel. \\
System & (30\,GiB) & (MB/tok)\tnote{b} & (MB/tok) & energy ($\mu$J/tok) & energy \\
\midrule
HBM-GPU (H20)        & 1 & 331 & 331 & 1326 & $20\times$ \\
\textbf{ReXpert} & \textbf{25} & \textbf{331} & \textbf{0}
  & \textbf{66} & \textbf{$1\times$} \\
\midrule
\multicolumn{2}{@{}l}{ReXpert without broadcast pooling\tnote{c}}
  & 420 & 0 & 84 & $1.3\times$ \\
\bottomrule
\end{tabular}
\begin{tablenotes}[flushleft]\footnotesize
\item[a] Dies needed to hold the $30.1$\,GiB of FFN expert weights
(Qwen3.5-35B-A3B-FP8).
\item[b] $L\cdot U(B)\cdot P_e/B$ with the measured union $U(64){=}168.6$ of $256$
experts: each activated expert is fetched once per step, by HBM grouping on the GPU
and by one broadcast ReRAM read-out on ReXpert. Energy counts weight movement only, at
$4.00$ (HBM) and $0.20$ (ReRAM) pJ/byte, so the ratio is exactly $E_{\text{HBM}}/
E_{\text{ReRAM}}$ at every batch.
\item[c] One read-out per four-token PE job instead of one per expert:
$213.5$ PE jobs vs $168.6$ union read-outs per layer. Pooling is a weight-traffic
mechanism, not only a latency one.
\end{tablenotes}
\end{threeparttable}
\end{table}

\begin{table}[t]
\centering
\caption{FFN-pool latency under equal-compute and equal-bandwidth
normalization.}
\label{tab:norm}
\begin{threeparttable}
\scriptsize
\setlength{\tabcolsep}{3.4pt}
\begin{tabular}{@{}llccccc@{}}
\toprule
& & & \multicolumn{2}{c}{Per-token ($\mu$s)} & \multicolumn{2}{c}{ReXpert faster by} \\
\cmidrule(lr){4-5}\cmidrule(lr){6-7}
Held equal & SKU & Dies & $B{=}8$ & $B{=}64$ & $B{=}8$ & $B{=}64$ \\
\midrule
\multirow{2}{*}{Peak compute}
  & H100/H800 & 1.6   & 145.8 & 61.25 & $75.6\times$ & $75.6\times$ \\
  & H20       & 10.8  & 18.26 & 7.67  & $9.5\times$  & $9.5\times$ \\
\midrule
\multirow{2}{*}{Read bandwidth}
  & H100/H800 & 122   & 1.93  & 0.81  & $1.0\times$ & $1.0\times$ \\
  & H20       & 102   & 1.93  & 0.81  & $1.0\times$ & $1.0\times$ \\
\midrule
\multicolumn{2}{@{}l}{\textbf{ReXpert}} & \textbf{25} & \textbf{1.93} & \textbf{0.81}
  & --- & --- \\
\bottomrule
\end{tabular}
\begin{tablenotes}[flushleft]\footnotesize
\item Equal-bandwidth rows match ReXpert at every reported batch because the measured
$U(B)$ keeps ReXpert read-bound through $B{=}64$; reaching our bandwidth takes
$102$--$122$ GPU dies against our $25$ (single-die equal-capacity pools are
$102$--$122\times$ slower).
\end{tablenotes}
\end{threeparttable}
\end{table}

\subsection{RQ2: MFU Cascade Ablations}
\label{sec:eval-mfu}
Table~\ref{tab:ablation} separates mechanisms under the factorization of
Eq.~\ref{eq:mfu}.
Broadcast, coactivation, and fetch move occupancy (and thus actual MFU) at
nearly fixed ideal MFU; gains are not additive across rows because each row
holds a different baseline configuration.
Broadcast is the primary occupancy lever at the side-4 point; placement and
fetch provide secondary recovery once the core mesh is fixed.
\begin{table}[t]
\centering
\caption{Per-mechanism ablations at the side-4 operating point.}
\label{tab:ablation}
\begin{threeparttable}
\scriptsize
\setlength{\tabcolsep}{2.6pt}
\renewcommand{\arraystretch}{1.05}
\begin{tabular}{@{}lcccc@{}}
\toprule
Mechanism (A$\to$B), config & Ideal MFU & Occupancy & Actual MFU & Rounds/str. \\
\midrule
\makecell[l]{Mask $w{=}4\to1$\tnote{a}\\ \enspace($B{=}8$, $G{=}16$)} & 0.346$\to$1.000 & 0.152$\to$0.199 & 0.050$\to$0.199 & 1.00$\to$1.01 \\
\makecell[l]{Bcast: direct$\to$bcast\\ \enspace(contiguous, $G{=}16$)} & 0.736$\to$0.736 & 0.152$\to$0.510 & 0.111$\to$0.374 & 8.89$\to$2.21 \\
\makecell[l]{Place: LPT$\to$coact\\ \enspace(bcast, $G{=}16$)} & 0.736$\to$0.736 & 0.519$\to$0.532 & 0.381$\to$0.391 & 2.16$\to$2.07 \\
\makecell[l]{Fetch: FIFO$\to$load-aware\tnote{b}\\ \enspace(bcast, $G{=}2$)} & 1.000$\to$0.778 & 0.422$\to$0.552 & 0.422$\to$0.420 & 2.45$\to$2.03 \\
\bottomrule
\end{tabular}
\begin{tablenotes}[flushleft]\footnotesize
\item The parenthesized configuration is held fixed while the mechanism is
toggled; actual MFU $=$ ideal MFU $\times$ occupancy (Eq.~\ref{eq:mfu}).
Default Qwen3.5-35B Alpaca trace.
\item[a] The width row uses roofline-adjusted ideal MFU
(\S\ref{sec:tilewidth}).
\item[b] Load-aware fetch admits narrower jobs to keep borrower PEs busy:
ideal MFU falls while occupancy rises; the net gain is the round-count drop.
\end{tablenotes}
\end{threeparttable}
\end{table}

\subsection{RQ3: Pooling Knee, Placement, Fetch, and Tile Width}
\label{sec:eval-balance}
Pooling has a bounded optimum (Table~\ref{tab:pooling}).
Side-$4$ raises Qwen3.5-35B occupancy $0.328\to0.519$ and actual MFU
$0.240\to0.381$ (ideal MFU stays at $0.736$), while larger domains raise
occupancy further but stretch rounds ($1.69$--$2.50\times$) so effective MFU
falls.
After BookSim2+DSENT network area, side~4 also minimizes the network area
spent per unit of delivered occupancy (mm$^2$/occ.\ in
Table~\ref{tab:pooling}); the measured Qwen3.5-397B trace reproduces the
same knee.
\begin{table}[t]
\centering
\caption{Core-pooling design-space sweep under measured routing.}
\label{tab:pooling}
\begin{threeparttable}
\scriptsize
\setlength{\tabcolsep}{2.6pt}
\renewcommand{\arraystretch}{1.05}
\begin{tabular}{@{}llcccccccc@{}}
\toprule
Model & Core & Ideal & \multicolumn{2}{c}{Occupancy} & Actual & Round & Effective & Net & mm$^2$/ \\
 & & MFU & bcast & direct & MFU & stretch & MFU & mm$^2$ & occ.\ $\downarrow$ \\
\midrule
\multirow{4}{*}{Qwen3.5}
 & $2{\times}2$ & 0.736 & 0.328 & 0.158 & 0.240 & $1.00\times$ & 0.240 & 26.0 & 11.96 \\
 & $\mathbf{4{\times}4}$ & \textbf{0.736} & \textbf{0.519} & \textbf{0.154} & \textbf{0.381}
   & $\mathbf{1.23\times}$ & \textbf{0.310} & \textbf{28.4} & \textbf{7.56} \\
 & $6{\times}6$ & 0.736 & 0.532 & 0.155 & 0.391 & $1.69\times$ & 0.231 & 30.6 & 10.88 \\
 & $8{\times}8$ & 0.736 & 0.678 & 0.182 & 0.499 & $2.50\times$ & 0.200 & 25.1 & 11.27 \\
\midrule
\multirow{4}{*}{Qwen3.5-397B}
 & $2{\times}2$ & 0.756 & 0.377 & 0.174 & 0.280 & $1.03\times$ & 0.273 & 26.0 & 4.42 \\
 & $\mathbf{4{\times}4}$ & \textbf{0.756} & \textbf{0.599} & \textbf{0.176} & \textbf{0.447}
   & $\mathbf{1.31\times}$ & \textbf{0.341} & \textbf{28.4} & \textbf{3.00} \\
 & $6{\times}6$ & 0.756 & 0.675 & 0.172 & 0.506 & $1.73\times$ & 0.293 & 30.6 & 3.29 \\
 & $8{\times}8$ & 0.756 & 0.803 & 0.212 & 0.603 & $2.02\times$ & 0.298 & 25.1 & 3.28 \\
\bottomrule
\end{tabular}
\begin{tablenotes}[flushleft]\footnotesize
\item Metrics per Eq.~\ref{eq:mfu}; effective MFU
$=\text{actual}/\text{stretch}$; mm$^2$/occ.\ is network area per delivered
occupancy; network area from BookSim2+DSENT.
Each model at its native tile-fill batch (35B: $B{=}128$; 397B: $B{=}205$).
\end{tablenotes}
\end{threeparttable}
\end{table}

Coactivation-aware placement provides secondary recovery
(Table~\ref{tab:ablation}: occupancy $0.519\to0.532$, rounds
$2.16\to2.07$); load-aware fetch helps mainly when few PE groups share a
core ($G{=}2$: $0.422\to0.552$ occupancy), since side-4 pooling at
$G{=}16$ already absorbs most residual imbalance.
Tile narrowing is a negative result (Table~\ref{tab:ablation}): at
$B{=}8$, $w{=}4\to1$ lifts ideal MFU $0.346\to1.000$ but the
roofline-adjusted peak falls to $0.256$, so $w{=}4$ remains the
resource-balanced choice.
The architecture therefore keeps the balanced tile and recovers occupancy by
sharing a read-out, rather than by shrinking the job.

\subsection{RQ4: AF-Disaggregated Serving with an H20 Attention Pool}
\label{sec:eval-afslo}\label{sec:eval-network}
We now assemble the AF system the architecture targets: H20 dies execute
attention (projections $+$ KV reads), ReXpert executes the MoE FFN, and the
two stages pipeline across decode waves---the polarity Raptor argues for MoE
serving \cite{nair2026raptor}.
Table~\ref{tab:afslo} compares it against a homogeneous H20 pool of the same
die count serving both stages.
On Qwen3.5-35B the AF split lowers decode TPOT by $1.25$--$4.0\times$ at
$4$K--$32$K context; the gain grows with model scale
($2.4$--$10.3\times$ on 397B, $2.5$--$10.4\times$ on
GLM-5.2) because the larger resident pools pair with proportionally larger
attention pools, whose KV reads then dominate less of the serial baseline.
When context grows the AF system itself becomes attention-bound: matching
stage times at $32$K needs ${\sim}7.5\times$ more attention dies than FFN dies
at $B{=}8$ on 35B (only $1.7\times$ on 397B, whose FFN pool is
larger).
The provisioning quantity is thus the attention/FFN die ratio
\cite{liu2026chime}.

\begin{table}[t]
\centering
\caption{AF-disaggregated serving (H20 attention $+$ ReXpert FFN) vs a
homogeneous H20 pool. All three models use measured serving traces.}
\label{tab:afslo}
\begin{threeparttable}
\scriptsize
\begin{tabular*}{\linewidth}{@{\extracolsep{\fill}}llcccccc@{}}
\toprule
Model & $B$ & Ctx & $t_{\mathrm{attn}}$ & $t_{\mathrm{FFN}}$ &
\multicolumn{2}{c}{TPOT ($\mu$s/tok)} & AF \\
\cmidrule{6-7}
 & & & ($\mu$s) & ($\mu$s) & AF & H20-only & gain \\
\midrule
\multirow{4}{*}{Qwen3.5-35B}
 & 8  & 4K  & 6.06 & 1.93 & 6.06 & 24.3 & $4.0\times$ \\
 & 8  & 32K & 33.2 & 1.93 & 33.2 & 51.5 & $1.55\times$ \\
 & 64 & 4K  & 4.15 & 0.81 & 4.15 & 11.8 & $2.85\times$ \\
 & 64 & 32K & 31.3 & 0.81 & 31.3 & 39.0 & $1.25\times$ \\
\midrule
\multirow{4}{*}{Qwen3.5-397B}
 & 8  & 4K  & 1.57 & 1.30 & 1.57 & 13.9 & $8.8\times$ \\
 & 8  & 32K & 5.10 & 1.30 & 5.10 & 17.4 & $3.4\times$ \\
 & 64 & 4K  & 0.64 & 0.63 & 0.64 & 6.56 & $10.3\times$ \\
 & 64 & 32K & 4.16 & 0.63 & 4.16 & 10.1 & $2.4\times$ \\
\midrule
\multirow{4}{*}{GLM-5.2}
 & 8  & 4K  & 2.30 & 2.07 & 2.30 & 21.9 & $9.5\times$ \\
 & 8  & 32K & 6.99 & 2.07 & 6.99 & 26.6 & $3.8\times$ \\
 & 64 & 4K  & 0.87 & 0.87 & 0.87 & 9.11 & $10.4\times$ \\
 & 64 & 32K & 5.56 & 0.87 & 5.56 & 13.8 & $2.5\times$ \\
\bottomrule
\end{tabular*}
\begin{tablenotes}[flushleft]\footnotesize
\item Attention pool: iso-compute H20 dies (10.8 / 125 / 235 for the three
models); FFN pool: resident ReXpert dies (25 / 289 / 543).
AF pipelines the stages, TPOT $=\max(t_{\mathrm{attn}},t_{\mathrm{FFN}})$;
the homogeneous baseline runs both on the attention dies,
TPOT $=t_{\mathrm{attn}}+t_{\mathrm{FFN,H20}}$.
FFN-only decode TPOT, excluding EP/TP transfer overlap.
Latencies are not comparable across models: the larger models amortize
per-token KV and expert traffic over $10$--$20\times$ more dies.
\end{tablenotes}
\end{threeparttable}
\end{table}

As in \S\ref{sec:pacom}, the provisioned fabric covers GLM's busiest Die-NoC
link and Qwen3.5's p99 multicast demand: the side-4 fabric is a
\emph{provisioning} result---large enough for measured p99 demand, small
enough that further core growth loses to round stretch and mesh silicon.
That budget is the $32$K-depth budget; \S\ref{sec:pacom-depth} shows that
halving depth to $16$K doubles Die-NoC and D2D demand on the models that
still map, so a shallower array is not a free latency win.

\subsection{RQ5: Area, Power, and Limitations}
\label{sec:eval-cost}\label{sec:limits}
Table~\ref{tab:overhead} closes the die budget.
The dominant area is the resident ReRAM array itself; the multicast Core mesh,
the price of bounded sharing, costs $35.6$\,mm$^2$ and $19$\,W under DSENT's
12\,nm-scaled parameters---silicon that buys $44\%$ more delivered occupancy
per mm$^2$ than the next core size (RQ3).
Modeled end-to-end over the FFN pool (RTL-synthesized PE, DSENT network, and
ReRAM read energy under measured activity, and UCIe-calibrated D2D
\cite{dassharma2022ucie,melek2025ucie}), a die draws $\approx$47\,W, so
the 25-die Qwen3.5-35B pool draws
$\approx$1.2\,kW against $4.3$\,kW TDP for the iso-compute 10.8-die H20 pool
\cite{nvidia2023h20}: $3.7\times$ lower power at $9.5\times$ lower latency,
or $\approx$35$\times$ lower FFN-pool energy per token---consistent with the
$20\times$ weight-movement-only bound of Table~\ref{tab:efficiency}.
\begin{table}[t]
\centering
\caption{ReXpert die area and power budget ($20{\times}20$\,mm, 12\,nm).}
\label{tab:overhead}
\begin{threeparttable}
\scriptsize
\setlength{\tabcolsep}{3.2pt}
\begin{tabular}{@{}lccl@{}}
\toprule
Component & Area (mm$^2$) & Power (W) & Basis \\
\midrule
ReRAM array (1T1R) & 268 & 3.3 & macro density\tnote{a}; $0.2$\,pJ/B read \\
PE $+$ scheduler   & 82  & 19.5 & RTL, 12\,nm synthesis\tnote{c} \\
Core mesh ($512$\,b) & 35.6\tnote{b} & 19.0 & BookSim2$+$DSENT, 12\,nm \\
Die NoC ($512$\,b)   & 2.2  & 1.8  & BookSim2$+$DSENT, 12\,nm \\
D2D $+$ misc IO      & 12 & 3.0 & UCIe PHY, 12\,nm-scaled\tnote{d} \cite{dassharma2022ucie} \\
\midrule
Die total & 400 (budget) & $\approx$47 & end-to-end FFN system model \\
\bottomrule
\end{tabular}
\begin{tablenotes}[flushleft]\footnotesize
\item[a] Same-node reported 1T1R macro density \cite{tsmc2022reram12nm}.
\item[b] DSENT router/link models with device parameters scaled to the
12\,nm node; the Core mesh fits the $50$\,mm$^2$ NoC/D2D budget.
\item[c] From 12\,nm RTL synthesis ($0.4$\,pJ/FLOP), at FFN MFU $0.38$.
\item[d] Provisioned off-die bandwidth (Table~\ref{tab:commbudget}:
$96{+}34{=}130$\,GB/s per die, both directions) at the UCIe
standard-package targets of $0.5$\,pJ/b and ${\sim}66$\,GB/s/mm$^2$ at
16\,GT/s \cite{dassharma2022ucie}, scaled to 12\,nm.
\end{tablenotes}
\end{threeparttable}
\end{table}

Two limitations remain beyond the modeled pool.
First, the design assumes a once-programmed, read-mostly ReRAM weight store
at reported same-node macro density \cite{tsmc2022reram12nm}.
Embedded ReRAM still faces device variation, retention, endurance, and
yield headroom relative to mature logic SRAM/DRAM; closing those gaps---and
the ECC, verify, and refresh policies they imply---is a process and circuit
prerequisite for a production resident FFN pool, not a scheduling detail.
Second, PaCom shows that activation and reduction traffic is small against
the provisioned hierarchy (Table~\ref{tab:commbudget}), but
\emph{when} and \emph{in what order} those messages share the Core mesh,
Die NoC, and D2D links under routing skew is still open:
contention-aware multicast arbitration, token-fetch versus compute
overlap, and cross-die reduction scheduling are left for further study.

\section{Related Work}
\label{sec:related}

\begin{table}[t]
\centering
\caption{Closest systems vs ReXpert (qualitative).}
\label{tab:relatedcmp}
\begin{threeparttable}
\scriptsize
\setlength{\tabcolsep}{1.6pt}
\begin{tabular}{@{}llll@{}}
\toprule
System & Locus & Skew/AFD handle & Diff.\ from us \\
\midrule
PUMA~\cite{ankit2019puma} & ReRAM CIM & --- & fixed tile--MAC bind \\
NeuPIMs~\cite{heo2024neupims} & DRAM-PIM+NPU & attn GEMV on PIM & FFN on NPU \\
PAPI~\cite{he2025papi} & GPU+hybrid PIM & dynamic FC offload & kernel steering \\
IANUS~\cite{seo2024ianus} & NPU--PIM & --- & area/power closure \\
CENT~\cite{gu2025cent} & CXL near-bank & e2e LLM & no fill pooling \\
Raptor~\cite{nair2026raptor} & 3D-DRAM & GPU-attn AFD & DRAM; no pooling \\
DIAMoND~\cite{liang2026diamond} & NAND+DRAM & expert select/place & edge, $B{=}1$ \\
CHIME~\cite{liu2026chime} & DIMM-PIM+GPU & \textbf{AFD} & attn pool only \\
EARTH~\cite{liu2026earth} & MoE accel. & prefetch/compr. & moves experts \\
Patterns~\cite{yu2026patterns} & multi-GPU & EP forecast & absorb in-core \\
\midrule
\textbf{ReXpert} & \textbf{ReRAM NMC} & \textbf{bounded mcast / FFN} & --- \\
\bottomrule
\end{tabular}
\end{threeparttable}
\end{table}

\textbf{ReRAM and PIM accelerators.}
ISAAC, PRIME, and PUMA establish input-stationary ReRAM CIM; CASCADE,
TIMELY, Ouroboros, and CompAir extend its movement and scaling
\cite{shafiee2016isaac,chi2016prime,ankit2019puma,chou2019cascade,li2020timely,liu2026ouroboros,li2026compair}.
NeuPIMs and IANUS pair GEMV PIM with GEMM NPUs
\cite{heo2024neupims,seo2024ianus}; CENT builds CXL near-bank PIM for
end-to-end LLM inference \cite{gu2025cent}; PAPI steers decode kernels
between GPU and hybrid PIM \cite{he2025papi}.
These designs bind compute to programmed weights, target attention-side
GEMV/KV traffic, or schedule kernels between fixed pools; ReXpert instead
uses digital NMC so PE groups remain reassignable under routing skew.

\textbf{MoE accelerators and traffic.}
EARTH predicts and prefetches compressed experts \cite{liu2026earth}; DIAMoND
places and selects experts across in-NAND/near-DRAM compute at the edge
\cite{liang2026diamond}; MoE-Lightning schedules GPU memory
hierarchies \cite{cao2025moelightning}; Patterns forecasts multi-unit expert
traffic \cite{yu2026patterns}; Tutel, FasterMoE, Lina, and SmartMoE balance
experts across GPUs
\cite{hwang2023tutel,he2022fastermoe,li2023lina,zhai2023smartmoe};
Expert Choice changes routing itself \cite{zhou2022expertchoice}.
We instead absorb skew inside a resident pool with routing fixed, neither
migrating, compressing, nor reselecting experts.

\textbf{Disaggregated serving.}
DistServe/Splitwise and Orca/vLLM/Mooncake establish disaggregated serving and
continuous batching
\cite{zhong2024distserve,patel2024splitwise,yu2022orca,kwon2023vllm,qin2025mooncake}.
CHIME's AFD$+$DIMM-PIM specializes the attention pool and quantifies
Attention--FC provisioning \cite{liu2026chime}; Raptor's 3D-DRAM silicon
argues the same GPU-attention/near-memory-FFN polarity, with expert weights
behind stacked bandwidth \cite{nair2026raptor}.
ReXpert complements both: ReRAM bandwidth density and bounded multicast
pooling on the MoE FFN side, with expert routing fixed
(Table~\ref{tab:relatedcmp}).

\section{Conclusion}
\label{sec:conclusion}
Disaggregated MoE decode asks the FFN pool for bandwidth density under
SLO-capped batches and sparse expert unions, then for occupancy recovery under
routing skew---without a global sharing fabric.
Residency alone is not enough: it removes off-chip weight delivery but exposes
tile underfill and straggler idle time.
ReXpert answers with a ReRAM near-memory FFN pool that factors actual MFU into
ideal MFU and occupancy, and recovers occupancy through bounded Core-local
broadcast, coactivation-aware placement, and load-aware fetch, while
provisioning each communication level to the induced demand.
Under the same measured$+$modeled hybrid analysis, side-4 pooling raises
occupancy $0.328{\to}0.519$, iso-peak-compute FFN-pool latency falls by
$9.5\times$ versus H20 at $20\times$ lower weight-movement energy, and an
H20-attention~$+$~ReXpert-FFN system lowers decode TPOT by
$1.25$--$4.0\times$, $2.4$--$10.3\times$, and $2.5$--$10.4\times$ on the
three models versus a homogeneous H20 pool.
The takeaway is architectural: for disaggregated MoE FFNs, bandwidth
density and bounded locality-preserving sharing matter more than batch
size alone.
ReRAM device maturity and contention-aware communication scheduling remain
open (\S\ref{sec:limits}).

\bibliographystyle{IEEEtran}
\bibliography{reference}

\end{document}